# Trends on 3d Transition Metal Coordination on Monolayer MoS$_2$

He Liu[1,†], Walner Costa Silva[2,†], Leonardo Santana Gonçalves de Souza[2], Amanda Garcez Veiga[2], Leandro Seixas[3], Kazunori Fujisawa[4,5,6], Ethan Kahn[7], Tianyi Zhang[7], Fu Zhang[7], Zhuohang Yu[7], Katherine Thompson[1], Yu Lei[7], Christiano J. S. de Matos[3], Maria Luiza M. Rocco[2], Mauricio Terrones[1,5,6,7,*], Daniel Grasseschi[2,*]

**Affiliations**
[1]Department of Chemistry, The Pennsylvania State University, University Park, PA, 16802.
[2]Institute of Chemistry, Federal University of Rio de Janeiro (UFRJ), 21941-909, Rio de Janeiro, Brazil.
[3]MackGraphe-Graphene and Nanomaterials Research Center, Mackenzie Presbyterian University, 01302-907, São Paulo, Brazil.
[4]Research Initiative for Supra-Materials (RISM), Shinshu University, 4-17-1 Wakasato, Nagano, 380-8553, Japan.
[5]Department of Physics, The Pennsylvania State University, University Park, PA, 16802.
[6]Center for 2-Dimensional and Layered Materials, The Pennsylvania State University, University Park, PA, 16802.
[7]Department of Materials Science and Engineering, The Pennsylvania State University, University Park, PA, 16802.
† - Equal contributors
* - Corresponding authors

**Abstract**

Two-dimensional materials (2DM) have attracted much interest due to their distinct optical, electronic, and catalytic properties. These properties can be by tuned a range of methods including substitutional doping or, as recently demonstrated, by surface functionalization with single atoms, increasing even further 2DM portfolio. Here we theoretically and experimentally describe the coordination reaction between MoS$_2$ monolayers with 3d transition metals (TMs), exploring the nature and the trend of MoS$_2$-TMs interaction. Density Functional Theory calculations, X-Ray Photoelectron Spectroscopy (XPS), and Photoluminescence (PL) point to the formation of MoS$_2$-TM coordination complexes, where the adsorption energy trend for 3d TM resembles the crystal-field (CF) stabilization energy for weak-field complexes. Pearson's theory for hard-soft acid-base and Ligand-field theory were applied to discuss the periodic trends on 3d TM coordination on the MoS$_2$ surface. We found that softer acids with higher ligand field stabilization energy, such as $Ni^{2+}$, tend to form bonds with more covalent character with $MoS_2$, which can be considered a soft base. On the other hand, harder acids, such as $Cr^{3+}$, tend to form bonds with more ionic character. Additionally, we studied the trends in charge transfer and doping observed in the XPS and PL results, where metals such as Ni led to an n-type of doping, while Cu functionalization results in p-type doping. Therefore, the formation of coordination complexes on TMD's surface is demonstrated to be a promising and effective way to control and to understand the nature of the single-atom functionalization of TMD.

## Introduction

In 2D materials (2DM), functionalization and doping were commonly accomplished through the induction of defects in the crystalline lattice or through the formation of substitutional solutions, where atoms with a greater or lesser number of electrons are placed as dopants for the creation of semiconductors of type n or p, or through the formation of covalent bonds between organic molecules and atoms at the edges of the materials.[1–4] However, these interactions tend to be unstable and/or significantly alter the crystalline structure of the 2D materials, drastically modifying their intrinsic properties and stability. It was recently proposed that the physical adsorption of organic molecules or metallic atoms as a functionalization method can be an effective and damage-free way of altering the properties of 2D materials.[5–7]

Lei *et al.* showed theoretically and experimentally that the presence of nonbonding electron pairs on the surface of InSe can be exploited by applying Lewis acid-base concepts.[8] Thus, simple acid-base reactions could achieve the formation of complexes with Lewis acids, such as $Ti^{4+}$, and 2D materials. Depending on the strength of the acid, it is possible to control the material's Fermi level and the band gap.[8] This opens a vast range of possible applications for these materials to fabricate photovoltaic devices. Ding et al. showed by DFT that the adsorption of 3d transition metal (TM) atoms on phosphorene surface is favorable, except for Zn.[9] Arqum et al. showed by DFT that transition metals coordinate at a "triangular" site on the surface of phosphorene, and the geometry resembles that of an octahedral coordination complex.[10] This concept can be expanded for other 2D materials, such as graphene and transition metal dichalcogenides (TMDs).[11]

Recently Gangwar et al. studied via DFT the interactions of first row Single Atom of Transition Metal (SA-TM) with $MoS_2$.[12] They found that hollow (the position at the center of the hexagons), and Mo sites had very close adsorption energies, both higher than the S sites. The trend observed follows the energy stabilization profile of crystal field (CF) for high-spin complexes, which should explain why the interaction with Zn is not favorable. They predicted that there is some coexistence of ionic and covalent bonding between the TM and $MoS_2$.[12] In this context, Hai et al. described theoretically and experimentally how the TM functionalization of $MoS_2$ could affect oxygen evolution reactions (OER) on its surface. The $MoS_2$-TM complexes were synthesized by hydrothermal treatment, and the results showed a monoatomic dispersion functionalization. They found that there is a relationship between TM-S and TM-O bond order with the OER activity.[13] Liu et al. developed a $MoS_2$-Co catalyst for hydrodeoxygenation. The catalyst was made by hydrothermal treatment, and the Co atoms were dispersed atomically and occupied the Mo, hollow, and S vacancy sites randomly.[14]

Although all these works successfully described the formation of $MoS_2$-TM complexes, the rationalization of fundamentals trends of SA-TM TMDs doping and the effects of the TM coordination sphere on the doping level are still missing. Furthermore, a more controllable functionalization with TM is still needed.[15] We recently showed that isolated Au atoms on the $MoS_2$ surface cause electrons to flow out of the 2D material, increasing the hole concentration on $MoS_2$.[16] DFT theoretical results revealed that changes in the TM coordination sphere could adjust the $MoS_2$-TM binding energy. For example, the energy of the $MoS_2$-Au bond is twice as high when one coordination site on the Au atom is filled by one S atom on the surface of $MoS_2$ and the other three by Cl anions than when no Cl is present. When Cl anions are present, the electron transfer from $MoS_2$ to Au is more efficient, leading to more effective p-type doping and better performance in a field-effect transistors device. Additionally, the electronic band dispersion of $MoS_2$-Au is drastically affected by the number of Cl atoms on the Au coordination sphere.[16]

Here, we expanded this vision to evaluate the functionalization of $MoS_2$ monolayer with several 3d transition metal ions (Cr, Mn, Co, Ni, Cu, and Zn). We focus on the $MoS_2$-TM interaction to evaluate the formation of coordination complexes on the surface and correlate their properties with classical coordination complexes, thus facilitating functionalization and controlling its optical and chemical properties with minimal damage to the crystal structure. First, the coordination trends of 3d TM were evaluated by DFT, considering the interaction between MoS$_2$ and single TM or TM-Cl$_3$ species. The theoretical trend was confirmed by X-Ray photoelectron spectroscopy (XPS), photoluminescence (PL), and reflection electron energy loss spectroscopy (REELS). Additionally, the samples were characterized by optical, scanning electron, and transmission electron microscopy, which confirmed the single atom nature of the functionalization with metals such as Ni. Finally, all observed trends are rationalized in terms of crystal-field stabilization energy, the absolute hardness of the 3d TM, and their reduction standard potentials related to the MoS$_2$ conduction and valence bands.

## Methods

**Synthesis of monolayer MoS$_2$** was carried out by a salt-assisted CVD method. A detailed description can be found in our previous publication. [16] The flakes were transferred to a new Si/SiO$_2$ substrate to remove the excess Na salt and MoO$_3$ used in the growth process.

**Functionalization of MoS$_2$.** The MoS$_2$-TM complex formation was performed by dipping the Si/SiO$_2$ substrate with the CVD MoS$_2$ into an ethanol solution of chromium acetate, manganese acetate, iron(III) chloride, cobalt chloride, nickel chloride, copper sulfate, or zinc chloride with concentrations between $1\times10^{-3}$ to $1\times10^{-9}$ mol L$^{-1}$ for 10 min. The functionalized M-MoS$_2$ sample was then immersed in isopropanol (IPA) for a few seconds to remove excess of transition metal salts, followed by N$_2$ drying. Finally, the sample was kept in a vacuum for 10 min before the measurements. Pristine samples were also washed with ethanol and IPA, dried with N$_2$, and kept in a vacuum for 10 minutes to exclude the effect of ethanol or IPA adsorption on the flake surface.

**Raman and Photoluminescence (PL) spectroscopy** were performed by excitation at 488 nm in a Microscope-based Renishaw INVia Spectrometer with thermoelectric CCD. For measurements with different transition metal concentrations, we followed approximately the same region of the same flake before and after functionalization. At least five different flakes were measured for each MoS$_2$-TM complex and TM concentration. PL mappings were obtained using an Alpha 300R Witec confocal Raman microscope with a highly sensitive EMS detector at 488 nm excitation wavelength and 2mW laser power.

**XPS and REELS measurements** were conducted in a high-resolution Thermo Scientific ESCALAB 250Xi spectrometer equipped with an electron energy hemispherical analyzer and using monochromatized Al Kα line (1486.6 eV) excitation. The spectra were energy referenced to the C1s signal of aliphatic C atoms at the binding energy of 284.8 eV. XPS spectra were collected using X-ray beam spot size = 650 μm with an emission angle of 90º with respect to the sample surface. High-resolution spectra were acquired with 25 eV pass energy. REELS spectra were acquired with the electron source operating at 1 keV.

**The computational approach** to study the equilibrium structure, stability, and electronic structure of transition metal doped $MoS_2$ is based on density functional theory (DFT) as implemented in SIESTA package. We used DZP localized basis, norm-conserved pseudopotentials with Troullier-Martins parametrization, mesh cutoff energy of 350 Ry, and k points sampling of Brillouin zone in Monkhorst-Pack algorithm with 10 x 10 x 1 grid. The exchange-correlation functional used is based on PBE generalized-gradient approximation. The calculations were performed within a supercell framework with 3 x 3 unit cells.

**Electron Microscopy imaging.** Scanning transmission electron microscopy (STEM) was carried out in an FEI Titan[3] G2 60/300 operated at 80 kV to reduce irradiation damage. A high-angle annular dark-field (HAADF) detector was used to collect the annular dark-field (ADF) signal. A Gaussian blur filter was applied using the *ImageJ* software to reduce the noise and enhance the visibility of the detailed structure, but raw images were used for acquiring the line profile of the ADF intensity. STEM-ADF image simulations were conducted using the QSTEM package. Simulation parameters such as acceleration voltage, spherical aberration ($C_3$ and $C_5$), convergence angle, and inner/outer angle for the HAADF detector were set according to experimental conditions. It is worth noting that prior to the STEM imaging, the CVD grown $MoS_2$ was first transferred to a TEM grid and then functionalized by $1\times10^{-6}$ mol $L^{-1}$ TM precursor in ethanol solution. Scanning electron microscopy (SEM) was carried out in an PhenomProX (ThermoFisher, Waltham, EUA), operated at 15 kV with backscattered electron detector.

# Results

**Prediction of $MoS_2$-TM complexes via DFT**

First, a screening simulation was done to evaluate the general trend of the coordination of different transition metals on the $MoS_2$ surface and correlate these trends with the classical coordination chemistry theory to rationalize the properties of chemically functionalized $MoS_2$-TM complexes. Figures 1a, b, and c shows the optimized structures for the semiconductor 1H phase of $MoS_2$ and the coordination of Co on two different coordination sites on the $MoS_2$ surface. There are three available coordination sites named H, M, and S sites (Figure 1a). In the M site, the coordinated metal is located on top of one Mo atom and bound to 3 S atoms, as shown in Figure 1b. In the H site, the TM is located over the center of the $MoS_2$ hexagonal structure and bound to 3 S atoms, as shown in Figure 1c. When the metal is coordinated in the S site, it is located directly on top of one sulfur atom and bound just to that atom, as is the case of Au atoms reported in a previous work of our group.[16] As we discussed in our previous work, [16] the TM coordination sphere significantly impacts their properties when bound to the $MoS_2$ surface. Thus, we considered two different cases in all calculations, in the first one, the TM is bound to the $MoS_2$ as a single adatom, named here $MoS_2$-TM. In the other case, the TM's coordination sphere was completed by Cl atoms forming an octahedral structure around the TM where it is bound to three S and three Cl atoms, labeled $MoS_2$-TM-$Cl_3$.

Figure 1d shows the $E_{ads}$ modulus as a function of the number of electrons in the 3d orbitals. $E_{ads}$ is defined as: $E_{ads} = E_{MoS2-TM} - (E_{TM} + E_{MoS2})$, where $E_{MoS2-TM}$ is the energy of the final $MoS_2$-TM complex, and $E_{TM} + E_{MoS2}$ are the energies of isolated TM and $MoS_2$, respectively. A clear trend is observed that resembles the one observed for the crystal field stabilization energy and the hydration enthalpy for weak field 3d TM complexes. The closed-shell ($d^{10}$) and the half shell ($d^5$) present the lowest adsorption energies due to the spherical symmetry of this electronic configuration. For other configurations, the non-symmetrical electronic distribution and the degeneracy loss of 3d orbitals lead to a stability gain upon the coordination, increasing the adsorption energy, as one can see, e.g., for Ni with $d^8$ electronic configuration.

To investigate the most stable coordination site, we plotted the adsorption energy difference between the H and M sites for all 3d TMs, as shown in Figure 1e. Here, negative values indicate that the M site is the most stable, and positive values means that the H is the most favorable coordination site. As one can see, the H site is the most favorable coordination site for $MoS_2$-TM, except for Sc and Cu that absorb at the M sites. On the other hand, for all the $MoS_2$-TM-$Cl_3$ complexes, there are only slight energy differences between the H and M sites, with the M site showing slightly higher $E_{ads}$ for all TMs, except for the Sc and Cu. This trend indicates that for all $MoS_2$-TM-$Cl_3$ complexes, at mild conditions, there are no preferable adsorption sites. For all 3d TMs the S were not stable and the structure relaxed with the TMs on H or M sites.

Figure 1f shows the TM's partial charge for $MoS_2$-TM and $MoS_2$-TM-$Cl_3$ complexes analyzed by the Voronoi deformation density method. We notice that TMs with a harder acid character (Sc, Ti, and V) tend to donate more electrons and stabilize with a higher positive charge, meaning that the metal-sulfur interaction exhibits some ionic character. This behavior is related to these metals' small ionization energy, leading to a greater tendency to lose the 4s and 3d electrons, leaving the atom with a closed shell and higher positive charges. In contrast, for softer metals, such as Co, Ni, and Cu, there is a tendency to form stronger covalent bonds, increase the charge delocalization, and decrease the metal's charge. Since sulfur atoms have a soft base character, the interaction with soft metals has a higher adsorption energy due to its more covalent character, as seen by the high $E_{ads}$ for Co and Ni. Furthermore, when the TM's coordination sphere is completed with Cl atoms, the partial charge is higher for all 3d metals.

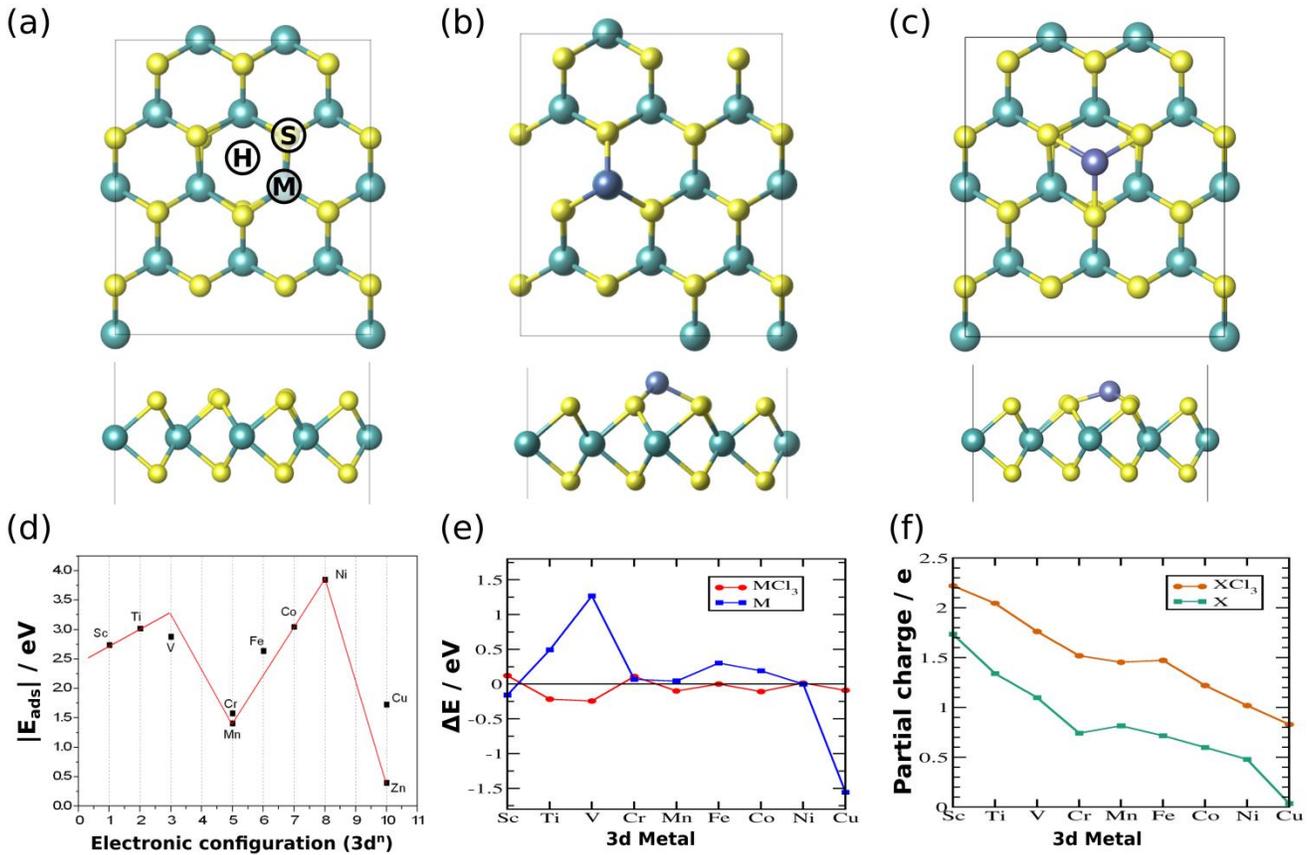

**Figure 1: DFT theoretical analyses of M-MoS$_2$ complexes.** Optimized (relaxed) structures for (a) pristine MoS$_2$ (possible coordination sites indicated); (b) MoS$_2$-Co in the M site configuration; and (c) MoS$_2$-Co in the H site configuration. (d) Adsorption energy as a function of the transition metal electronic configuration for MoS$_2$-TM complexes. (e) H- and M-site adsorption energy difference for the 3d TMs for MoS$_2$-TM and MoS$_2$-TM-Cl$_3$ coordination compounds. (f) Partial charge for the TMs in the cases of MoS$_2$-TM and MoS$_2$-TM-Cl$_3$ coordination compounds.

The electronic band dispersion was calculated to investigate the influence of TM and TM-Cl$_3$ coordination on the electronic properties of MoS$_2$. Figure 2 shows the band dispersion for MoS$_2$-Co-Cl$_3$, MoS$_2$-Ni-Cl$_3$, and MoS$_2$-Cu-Cl$_3$ on both H and M sites as well as for pristine MoS$_2$. We noticed that for all calculated 3d TMs, the material remained with a direct gap after functionalization. However, the Fermi level energy, the states near the Fermi level, and the spin polarization of these states were drastically affected by the TM's coordinated at the MoS$_2$ surface (Figure 2). For example, the adsorption of Co atoms creates new states near the Fermi level, between 0 and 0.5 eV, localized on the Co atoms when it is coordinated on the H site (Figure 2b). On the other hand, when CoCl$_3$ is in the M site (Figure 2c), the new states are near the Fermi level and the conduction band (CB), positioned ~0.25eV under the CB. MoS$_2$→Co charge transfer transitions in the visible and near-infrared spectral range could access these new levels. In the case of NiCl$_3$ (Figure 2c and Figure 2f), the new levels localized above the Fermi level, at ~0.1 and ~0.75 and for both coordination sites and could be accessed by $MoS_2 \rightarrow Ni$ charge transfer transition in the mid and far-infrared region. The coordination of Cu atoms creates localized states above the Fermi level and near the valence band (VB) (Figure 2d and Figure 2g) at ~0.05 eV and ~0.5 eV for the coordination at the M and H site, respectively.

For the $MoS_2$-TM complexes the coordination of single TM atoms leads to a different band dispersion, e.g., for Co and Ni the new states are above the Fermi level, between -1 and 0 eV, and for Co they have different spin polarization (Figure S1). Therefore, the coordination of TM or TM-$Cl_3$ on the $MoS_2$ surface can be explored to carefully tune its optical, electronic, and chemical properties by choosing the TM and its coordination sphere. Another important aspect is that these controlled changes can be performed with minimal changes on the crystalline structure since there is no significant alteration of the Mo-S bond length and Mo-S-Mo bond angle after TM functionalization. This is a clear advantage compared to other types of doping strategies, such as controlling the defects density or interstitial/substitutional doping where high doping levels cannot be achieved without compromising materials stability and mechanical properties.[17] This way, based on the DFT calculations, we performed the functionalization of $MoS_2$ samples with 3d TM such as Cr, Mn, Co, Ni, Cu, and Zn to correlate the trends on $E_{ads}$ calculated by DFT, with the photoluminescence, XPS and REEL spectra as we shall see in following sections.

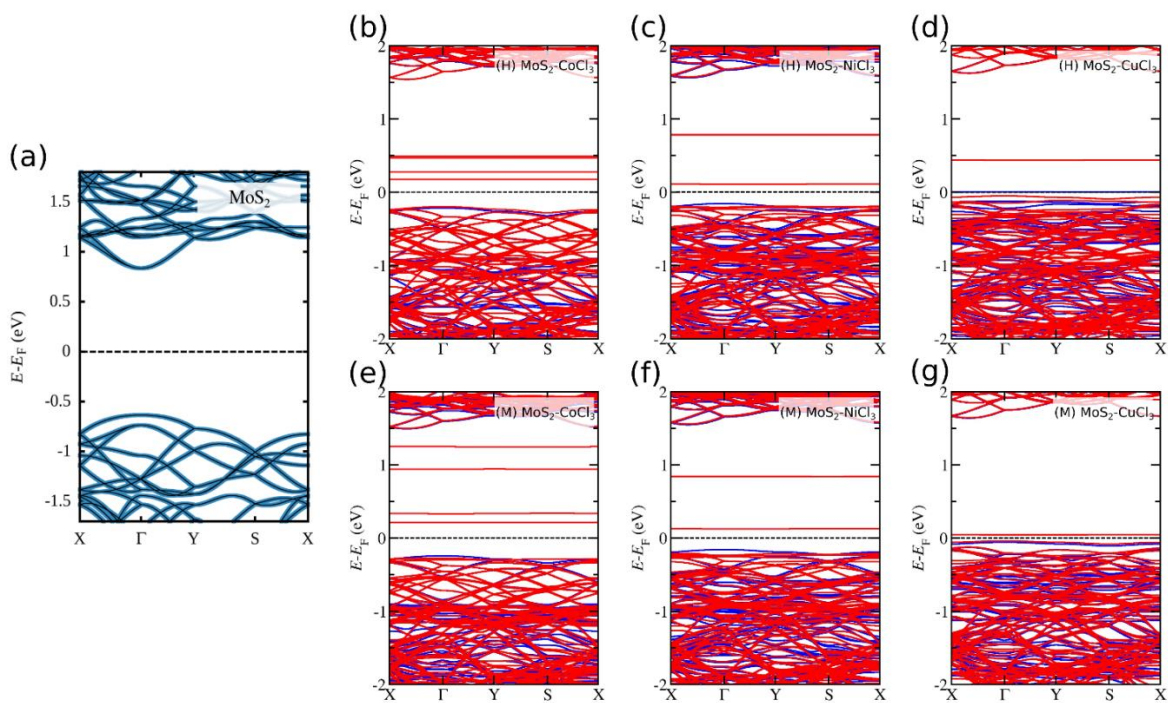

**Figure 2: Electronic Band Dispersion of TM functionalized MoS$_2$.** (a) Electronic structure of pristine MoS$_2$ calculated by DFT. (b-g) Electronic band dispersion of Co, Ni and Cu functionalization on MoS$_2$ via H (b-d) and M (e-g) sites with Cl$_3$ completing the TM's coordination sphere. The blue and red curves indicate states with different spin polarization. The "flat" states are new states localized on the TM atoms created after functionalization, indicating the possibility to control the doping type by chance the TM coordination site.

**Experimental characterization of MoS$_2$-TM coordination complexes**

**Electronic Doping of MoS$_2$ via TM functionalization.** TM coordination on MoS$_2$ was performed by exploring a straightforward acid-base reaction between a Lewis acid, the TM precursor, and a Lewis base, the MoS$_2$. During the reaction, sulfur's 3p valence electrons are donated to TM's 3d valence orbital, forming a MoS$_2$-TM coordination bond. The functionalizations were performed on CVD grown MoS$_2$ monolayers as described in the methods section. TM chlorides, acetates, and sulfates were chosen since they form labile complexes with TM in aqueous and ethanol solutions, favoring ligand exchange reactions on TM's coordination sphere. XPS measurements were conducted on MoS$_2$-TM samples functionalized in a TM solution of $1\times10^{-4}$ mol L$^{-1}$ to confirm the TM presence on the MoS$_2$ surface. Figure 3 depicts the XPS results and indicates that TM functionalization was effective since we observed shifts on Mo 3d and S 2p core-levels. The high-resolution spectra in specific regions of each TM (Figure S1) revealed the presence of the studied metals, and the chemical shift for each one can be related to its chemical environment and the chemical state.

To elucidate each TM's oxidation state, the core-level spectra were compared with known species referent to each TM, as shown in Figure S2. For Cr, its 2p core-level spectra are comparable to that of $Cr_2O_3$.[18] Cr 2p orbit presents a multiplet splitting with a binding energy of 576.7 eV for $Cr^{3+}$ $2p_{3/2}$ and 586.5 eV for $Cr^{3+}$ $2p_{1/2}$ and peak separation of 9.8 eV (Figure S2a), indicating the presence of $Cr^{3+}$ on the $MoS_2$ surface. For Mn, the presence of a satellite peak indicates an $Mn^{+2}$ oxidation state, and the spectra are comparable to that of MnO, with peaks at 641.2 and 653.2 eV for Mn $2p_{3/2}$ and $2p_{1/2}$, with a slight redshift of 0.2 eV (Figure S2b).[18] Co $2p_{3/2}$ 781.1 and $2p_{1/2}$ 797.1 eV peaks presented a blueshift of 1.4 eV and are comparable to that of CoO, with a spin-orbit splitting of 16.0 eV, and characteristics satellites, confirming the presence of $Co^{2+}$ (Figure S2c).[18] For Ni, we observed peaks at 856.0 and 873.6 eV, assigned to $2p_{3/2}$ and $2p_{1/2}$, receptively. The Ni 2p core-level spectra are comparable to $Ni(OH)_2$, with prominent blueshift and spin-orbit splitting of 17.6 eV, although Ni $2p_{3/2}$ peak seems asymmetric and comparable to NiO, indicating a mixture of species, with $Ni^{2+}$ state (Figure S2d).[18] For Cu, the XPS spectra shows a mixture of $Cu^{2+}$ and $Cu^{1+}$, since there is a satellite peak (938.0 to 947.1 eV) between the Cu $2p_{1/2}$ (952.7 eV) and $2p_{3/2}$ (932.9 eV) peaks, although this peak is broad and may have contributions of both $Cu^{2+}$ and $Cu^{1+}$ species (Figure S2e).[18] Thus, Cu 2p core-level XPS spectra indicate a reduction of $Cu^{2+}$ to $Cu^{1+}$ during the $MoS_2$ functionalization. Further results indicate that this reduction occurs spontaneously on the $MoS_2$ surface (see below for details). Zn spectrum has peaks at 1022.4 and 1045.4 eV, which are assigned to Zn $2p_{3/2}$ and $2p_{1/2}$, indicating a $Zn^{2+}$ oxidation state. It is in good agreement with that of ZnO, showing no shifts (Figure S2f).[18]

For all TMs studied here, their oxidation states remained unchanged after the functionalization, except for Cu, where a reduction in the oxidation state was observed. The shifts observed on the TM XPS spectra can be explained by the charge transfers between the TM and the $MoS_2$. Figure 3a shows the spectra of Mo 3d core-level of pristine and doped $MoS_2$. The deconvolution of pristine $MoS_2$ spectrum (Figure 3c) shows the characteristic $Mo^{4+}$ $3d_{5/2}$ peak at 230.0 eV that will reference for further discussion. When the $MoS_2$ is functionalized with a TM, charge transfer may occur, leading to new charge densities on Mo and TM atoms; therefore, a shift in $Mo^{4+}$ $3d_{5/2}$ peak is expected. In this context, a blueshift indicates a charge transfer from Mo to the TM, leading to a more negative character on the TM and a more positive character (p-type doping) on $MoS_2$, increasing the binding energy of Mo electrons. Moreover, a redshift indicates a charge transference from TM to

Mo, leading to a more positive TM and a more negative character (n-type doping) in $MoS_2$, decreasing the binding energy of its electrons.

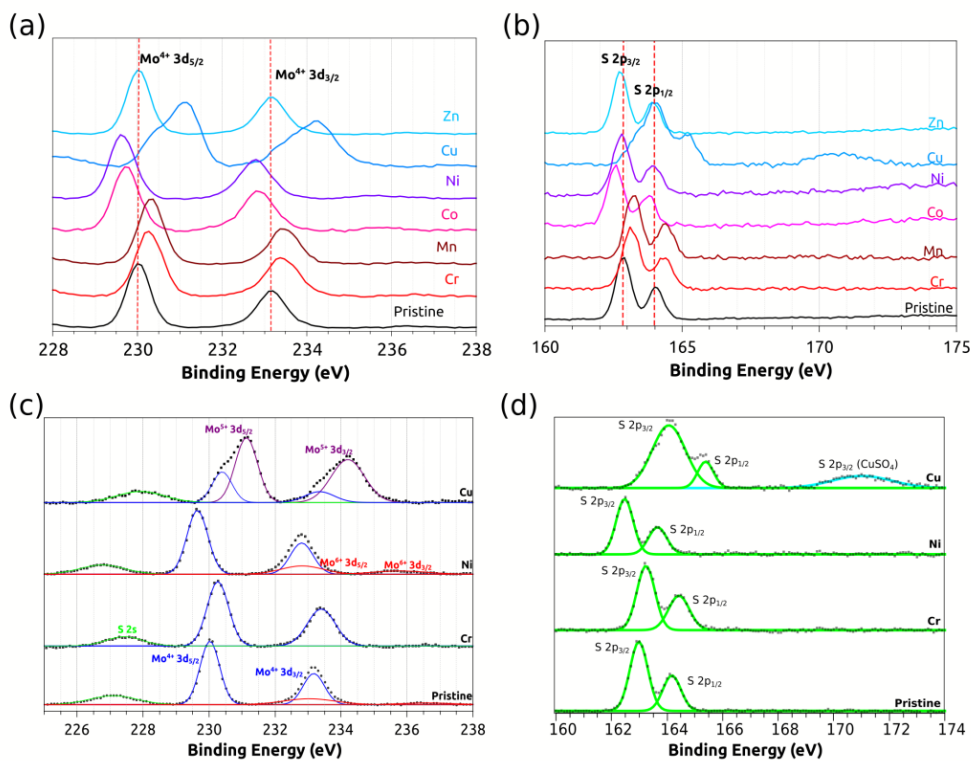

**Figure 3: X-ray photoelectron spectroscopy studies of MoS$_2$-TM complexes.** (a) XPS spectra of Mo 3d orbit on pristine and MoS$_2$-TM complexes. The curves are the original spectrum after C1s (284.8 eV) calibration. (b) S 2p orbit of pristine and MoS$_2$-TM, showing the S $2p_{1/2}$ and $2p_{3/2}$ peaks. (c) Deconvolution of Mo 3d spectra for pristine, MoS$_2$-Cr, MoS$_2$-Ni, and MoS$_2$-TM. The blue curves are the Mo$^{4+}$ $3d_{5/2}$ and $3d_{3/2}$, purple curves are the Mo$^{5+}$ $3d_{5/2}$ and $3d_{3/2}$ peaks and the red curves are the Mo$^{6+}$ $3d_{5/2}$ and $3d_{3/2}$ peaks. The green curves are the S 2s peak. (d) Deconvolution of S 2p spectra for pristine, MoS$_2$-Cr, MoS$_2$-Ni, and MoS$_2$-TM. The green curves are the S $2p_{1/2}$ and $2p_{3/2}$ peaks, and the blue curve is the S $2p_{3/2}$ for the CuSO$_4$ precursor.

The Mo 3d core-level spectra for the MoS$_2$-Cu samples indicate the oxidation of Mo$^{4+}$ to higher oxidation states. The deconvolution of Mo 3d spectra in Figure 3c points to the presence of Mo$^{5+}$ peaks at 231.1 and 234.2 eV. Thus, there is a spontaneous oxidation of Mo$^{4+}$ to Mo$^{5+}$ followed by the Cu$^{2+}$ to Cu$^{1+}$ reduction, as indicated by the Cu 2p spectra (Figure S1e). In Figure 3c, the deconvolution of Mo 3d spectra of pristine, MoS$_2$-Cr, and MoS$_2$-Ni samples are also presented, showing that the oxidation of Mo$^{4+}$ is observed only for Cu samples. The presence of Mo$^{6+}$ in some pristine and functionalized samples is originated from the synthesis.[19] The S 2p peak in the MoS$_2$-Cu samples are shifted to higher energies due to the change on Mo oxidation state, however it still in the typical range of sulfides.[18] For all the S spectra (Figure 3b and 3d), similar behavior is observed since $MoS_2$ has a high degree of mixing of Mo and S atomic orbitals.[20]

**Band gap tuning through TM coordination.**

$MoS_2$ is a layered semiconductor that shows the transition from indirect to direct band gap when it reaches a monolayer.[21] To evaluate if the band gap remains direct after the TM functionalization and possible shifts on the band gap values, REELS measurements were performed. Figure 4 shows the Tauc-Plot obtained from the REELS spectra, where we can observe that the doping does not change the direct band gap character, indicated by the linear regression on the gap transition presented. All linear regression showed an $R^2$ of 0.99. Analyzing the interception between the linear regression and the baseline, small shifts on the band gap value can also be observed (Figure 4). As we can see, in general, TM-functionalization, except Mn and Zn, lowered the bandgap of $MoS_2$, which indicates the creation of new states between VB and CB. However, due to the low energy resolution, only the shift observed for Cr and Co samples can be considered. Therefore, more experiments are needed to fully understand the influence of TM doping on $MoS_2$ band gap values. It is worthy to note that the shoulder between 1.6 and 2.4 eV can be assigned $MoS_2$ excitons and trions.[22] However, due to our energy resolution, they cannot be separated, and the TM's effects on the excitons will be discussed on the next section based on PL measurements.

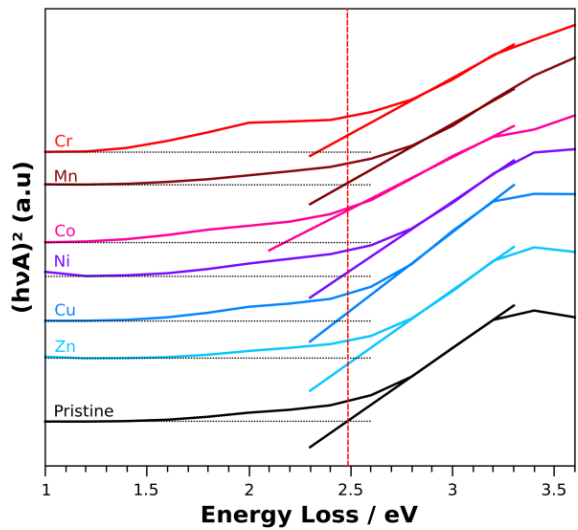

**Figure 4.** Tauc-Plot from the REELs data for $MoS_2$-TM indicating the band gap by interception between the linear regression and the baseline. All linear regressions presented $R^2=0.99$.

**Photoluminescence tuning through TM coordination.**

Optically generated electron-hole pairs in monolayer $MoS_2$ form stable exciton states even at room temperature because of the extremely large Coulomb interactions in atomically thin 2D materials. The neutral exciton plays an important role in the optical properties of the monolayer. Figure 5a shows the PL spectra for a typical monolayer with the main excitonic peak at 1.83 eV (exciton A) related to the direct band gap transition at K point and a small shoulder at 2.0 eV (exciton B) related to the transition at K' point with its energy being related to the level splinting due spin-orbit coupling. PL spectroscopy measurements were carried out for $MoS_2$-TM complexes at various TM concentrations ranging from $1 \times 10^{-9}$ mol $L^{-1}$ to $1 \times 10^{-6}$ mol $L^{-1}$ to study the effect of the TM coordination functionalizations on the optical properties of $MoS_2$. Figure 5a shows the PL spectra of pristine and Ni functionalized $MoS_2$ monolayers. The PL intensity and shape of the A exciton are significantly altered after Ni functionalization and changes as a function of TM concentration.

The PL peak can be deconvoluted and fitted into two Lorentzians, the neutral exciton (X) and the trion ($X^-$) to understand these spectral changes. It can also be seen that the $X^-$ peak was enhanced while the X intensity was quenched after the Ni functionalization. The formation of the coordination π backbonds between the S and Ni atoms results in Ni electrons being transferred to the $MoS_2$, leading to an increase in the trion intensity relative to the neutral exciton intensity. These changes on exciton and trion populations result in an overall PL quenching and widening due to MoS$_2$ functionalization, agreeing well with the XPS and DFT results aforementioned.[22] The exact opposite effect is seen for Cu functionalization, as shown in Figure 5e. After Cu functionalization, we note an increase in exciton intensity relative to the trion intensity, similar to Au functionalization, resulting from p-type doping of MoS$_2$.[16] Figure 5(b-d,f-h) further shows the exciton to trion ratio for other MoS$_2$-TM complexes. After TM functionalization, some metal ions such as Cr, Co, and Ni significantly quench the neutral exciton and enhance relative to the trion intensity. For Mn and Zn, we observe only a small quenching effect in the PL, which can be explained by their half shell and full shell 3d configuration.

To further verify the trends in PL spectra as a function of functionalization with the different metals, we plotted the normalized exciton to trion ratio percentage change as function 3d metal concentration, as shown in Figure S6. Like that induced by Ni coordination, a high decrease in the exciton-trion ratio was also observed for Cr and Co doping. Since the trion has a smaller PL efficiency, the overall PL intensity decreases with the Ni, Cr and Co doping. In contrast, for low concentrations ($1.10^{-9}$ mol L$^{-1}$ - $1.10^{-7}$ mol L$^{-1}$) of Zn, practically no PL changes were observed. We notice just a small intensity (less than 20% change) decrease for Zn and Mn functionalizations, probably due to salt deposition on the surface. These results show a remarkable resemblance to the trend of theoretical adsorption energy shown in Figure 1. As expected, Mn and Zn showed the smallest effect due to the half and closed-shell configuration, respectively, this leading to lower stabilization energy for the 3d orbitals upon their coordination to the sulfur atoms on the surface.

In contrast, Ni and Co showed the highest effect on the PL spectra, both in intensity and exciton/trion populations. These results suggest that the effects of Ni coordination on MoS$_2$ were maximized by the formation of a stronger covalent Ni-S bond, which increases the charge delocalization and the possibility of ligand-metal charge transfers. It is also worth noting that the change in PL is of positive correlation between the concentration of TM ions in all cases, regardless of n or p doping types.

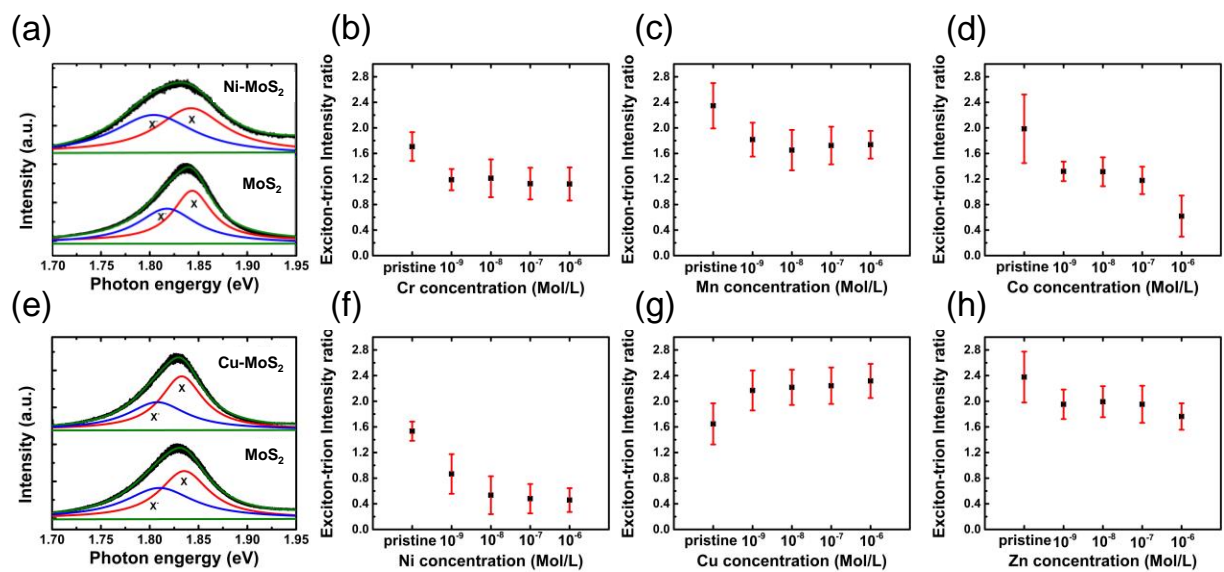

**Figure 5: PL spectra analysis of TM functionalized MoS$_2$.** (a) Photoluminescence spectrum of pristine and functionalized MoS$_2$ monolayers. The A exciton is deconvoluted into the trion (X-) (blue curve) and exciton (X) (red curve) peaks through Lorentzian functions. After functionalization, the trion intensity increased while the exciton intensity decreased, which corresponds well with the n-type doping effect of the Ni functionalization. (b) Photoluminescence spectrum of pristine and Cu functionalized MoS$_2$ monolayers. After functionalization, the trion intensity decreased while the exciton intensity increased, which corresponds well with the p-type doping effect of the Cu functionalization. (b-d and f-h) Exciton to trion intensity ratio of pristine and functionalized MoS$_2$ for different TM precursors: Cr (b), Mn (c), Co (d), Ni (f), Cu (g), and Zn (h) at different concentrations.

## Morphological Characterization

In order to characterize and identify the structure of MoS$_2$-TM complexes, morphological characterization was performed by optical microscopy, SEM, and HAADF-STEM. Figure 6a shows an optical image of the MoS$_2$-Ni sample where triangular flakes typical of monolayer MoS$_2$ can be observed. We also notice some multilayer regions with higher contrast and some black points of MoO$_3$ residue. Even though we did not observe the formation of TM with zero oxidation state by XPS, SEM was used to evaluate the possible formation of TM-nanoparticles on the MoS$_2$ surface.[13] Figure 6b depicts an SEM image of the MoS$_2$-Ni sample after functionalization with NiCl$_2$ solution at 10$^{-4}$ mol L$^{-1}$. The MoS$_2$ monolayer is highlighted in yellow, and as can be seen, even with a high TM concentration, no nanoparticles were present on the flake surface, corroborating with XPS results. High-resolution HAADF-STEM was conducted to study the atomic structure of the TM atoms on MoS$_2$. The ADF intensity changes depend on the Z-number of atoms ($\sim Z^{1.6-1.9}$),[23] thus, Ni atoms stand out in the MoS$_2$ lattice. As shown in Figure 3c, bright single Ni atoms on top of the MoS$_2$ lattice can be observed, similar to our previously published results on AuCl$_x$ functionalized MoS$_2$.[16] The imaging of other TMs studied, such as Co and Cr, were also performed, but due to their lower Z number and |E$_{ads}$| compared to Ni, we could not identify these atoms on top of the MoS$_2$ lattice.

It is worthy to note that the HAADF-STEM image reveals that Ni atoms were found on MoS$_2$ basal plane and not on the defect or edge regions, indicating that the Ni functionalization occurs by its coordination on the S atoms. To further evaluate TM coordination on the MoS$_2$ basal plane, PL mapping was performed on pristine and MoS$_2$-TM samples. The results for a MoS$_2$-Ni sample functionalized with Ni$^{2+}$ solution at 10$^{-7}$ mol L$^{-1}$ are shown in Figure S3. As discussed in the previous sections, Ni functionalized MoS$_2$ has decreased exciton/trion ratio and overall PL intensity since the trion has a smaller PL efficiency. Figure S2 shows that the intensity change is uniform on the whole flake, which indicates that Ni coordination is homogeneous on the MoS$_2$ basal plane. Similarly, we observed a redshift of the PL peak due to the increase in the trion population, which was more pronounced on the flake center. Figure S4 shows Co functionalization presenting similar phenomenon, indicating that the TM coordination is homogeneous on the entire flake.

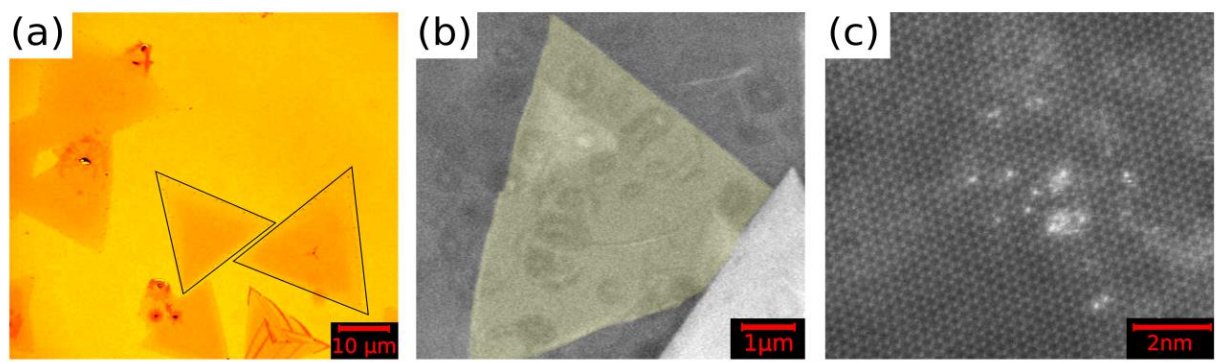

**Figure 6. Imaging of MoS$_2$ by different microscopy techniques.** (a) Optical Microscopy showing triangular MoS$_2$ monolayer flakes and dark spots of MoO$_3$. (b) SEM image of a Ni functionalized MoS$_2$ flake showing the absence of aggregates or Ni nanoparticles. (c) HAADF-STEM image of MoS$_2$-Ni, where the bright spots show Ni single atoms atomically dispersed on the MoS$_2$ basal plane.

## Discussion

Recently it was proposed that Classical Coordination Chemistry concepts could be applied to understand the single-atom functionalization of graphene and its derivatives.[11] In our previous work, we expanded this vision to explore the functionalization of MoS$_2$ with single Au atoms and its effect on a field-effect transistor.[16] Thus, here, coordination chemistry will be applied to understand the trends observed for the coordination of 3d-TMs on the MoS$_2$ surface. The coordination of SA-TM on the 2DM surface can be understood as forming coordination complexes on the material's surface.[11] Thus, a parallel between the molecular orbital theory for coordination complexes and 2DM solid-state physics can elucidate the trends and properties observed after the functionalization. The application of coordination chemistry and hard-soft acid-base concepts[16,24,25] can lead to greater understanding and control of the chemical, electric, and optical properties of 2D materials.

Classical Werner coordination complexes with closed-shell (d$^{10}$) and the half-shell (d$^5$) TM in a weak-field (low-spin) configuration do not present ligand-field stabilization energy (LFSE) due to the spherical symmetry of these electronic configurations.[26] In contrast, in TM with d$^3$ and d$^8$ electronic configurations, the low energy ligand orbitals have a higher electron population and consequently higher LFSE.[26] Our DFT calculations showed that TM coordination on M and H site have very close energies when the TM's coordination sphere is complete with Cl ligands. The TM will acquire an octahedral geometry similar to *fac* isomers of TMA$_3$B$_3$ complexes in both sites, where 3 S atoms are on one triangular face and the Cl atoms on the other face. Both S and Cl atoms are π donor ligands; thus, the TM should adopt a weak-field configuration. Since the MoS$_2$-TM functionalization was done in a very mild condition, we expect that TM's coordination sphere will be composed of three S atoms and three ligands from the precursor salt, chloride, or acetate in our case. Additionally, we expect that TM coordination on MoS$_2$ will follow the same trend observed for the LFSE showed in Figure 7a.

This way, MoS$_2$-TM complexes with d$^{10}$ (Zn) and d$^5$ (Mn and Cr) should have smaller adsorption energy (Figure 1d). For different configurations, the non-symmetrical electronic distribution and the degeneracy loss of 3d orbitals lead to a stability gain upon the coordination, increasing the adsorption energy. As for classical coordination complexes, Ni, with d$^8$ electronic configuration, shows the higher adsorption energy due to the complete occupancy of the low energy ligand orbitals.[26] This higher adsorption energy could explain that the MoS$_2$-Ni complex was the only one that is stable during e-beam irradiation on HAADF-STEM imaging, the more significant BE shift in XPS spectra, and the more pronounced decrease of exciton/trion ratio.

To better visualize the observed trends on XPS, REELS, and PL, we plotted the difference between the pristine and MoS$_2$-TM samples for each experimental parameter analyzed as a function of the number of electrons on d orbitals (n). Figure 7 compares all trends with the ligand-field stabilization energy for classical coordination complexes. Figure 7b shows the binding energy shift of Mo and S core-levels for $MoS_2$-TM samples. We can observe a blueshift for Cr, Mn and Cu, increasing the binding energy of $Mo^{4+}$ 3d electrons. For Co and Ni, a redshift was observed with ΔBE of -0.3 and -0.4 eV, respectively. For Zn, no considerable shift was observed on Mo 3d peaks. The S 2p core-level spectra indicate the same behavior observed in the Mo 3d spectra for all 3d TM studied here. This is expected since the VB's top, and the CB's bottom in the $MoS_2$ show a high degree of mixing of Mo and S atomic orbitals.[20] Therefore, both S and Mo core-level spectra should present similar shifts upon TM coordination. The higher shift observed for $MoS_2$-Ni samples can again be related to $Ni^{2+}$ higher ligand-field stabilization energy.

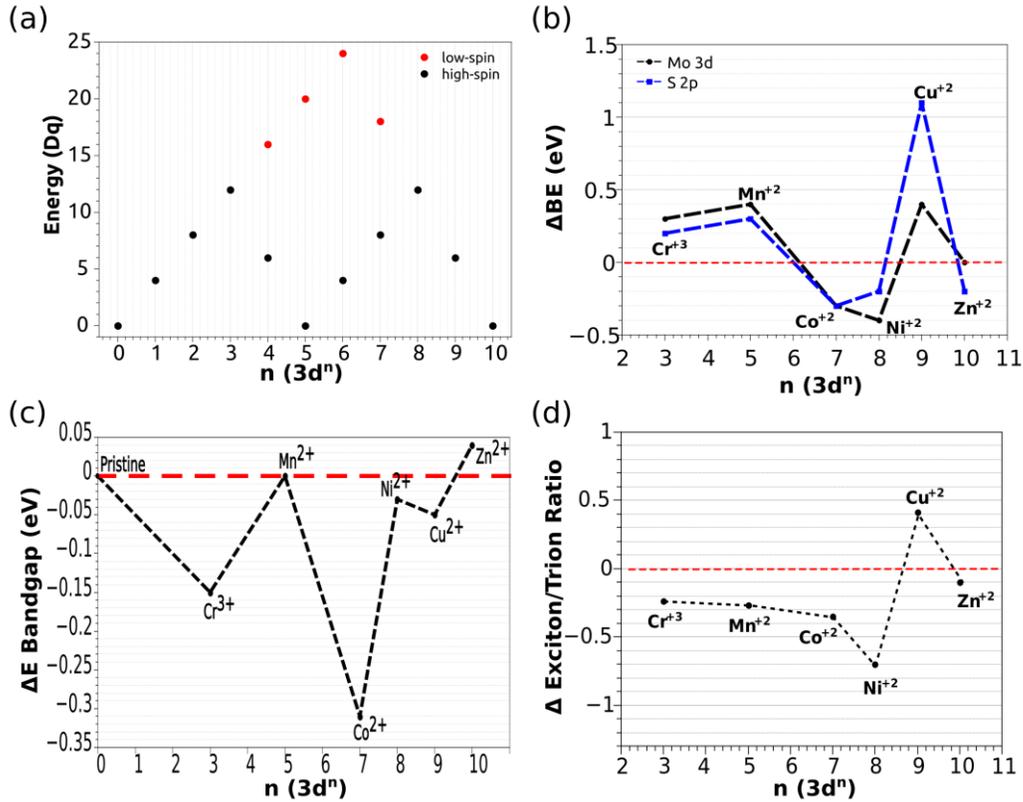

**Figure 7:** (a) Ligand-field Stabilization Energy for 3d TM complexes, showing the weak and strong-field configuration. The ligand-field splitting parameter Dq is equal to $Ze^2 r^4 = 6a^5$, where Z is the TM charge, e the elementary electron charge, r the TM radius, and the TM-ligand bond distance. (b) Binding energy shift for S $2p_{3/2}$ (blue curve) and Mo $3d_{5/2}$ peak as function of TM 3d electronic distribution for MoS$_2$-TM samples. (c) Band gap shift calculated from the Tauc-plots showed in Figure 4. (d) Normalized exciton to trion intensity changes showing the percent change in exciton to trion ratio for various TM functionalized MoS$_2$ samples at different TM concentrations. The TMs here can be separated into three groups. One is Cu, which shows an increase in the exciton to trion ratio, indicating p-type doping on the MoS$_2$. The second group is Zn, Mn, and Cr, where we see a small decrease in the ratio, indicating weak n-type doping of MoS$_2$. The last group is Co and Ni, where we see a large decrease in the ratio that corresponds to strong n-type doping of MoS$_2$.

The Pearson acid-base hard-soft theory can be applied to rationalize the direction of the BE energy shift. When the acid's LUMO level is closest to the base's HOMO level, the acid-base coordination bond has a more covalent character, and charge transference may occur between the species' frontier orbitals. On the other hand, when the acid and base valence states have a significant energy difference, the interaction between them has a more ionic character, and some electron withdrawal effects can be observed due to Coulomb interactions. In the case of MoS$_2$-TM complexes, we can estimate the TM LUMO by its electron affinity, and the base's HOMO level is the top of MoS$_2$ VB. This way, to elucidate the observed trends on XPS spectra of MoS$_2$-TM samples and how each TM interact with MoS$_2$, the energy level diagram in Figure 8 was created using the HOMO and LUMO orbitals of each TM, estimated by the ionization potential and electron affinity of the neutral TM, the $TM^{2+}/TM^{1+}$ redox potentials,[27] and the absolute energies of $MoS_2$ VB and CB.[28] Although the HOMO and LUMO TM's energy levels are referenced to the +1 oxidation state, and the XPS results indicate +2 oxidation states, we expected these energy levels to be lower; however, the same trend is expected, and it is comparable to the experimental data.

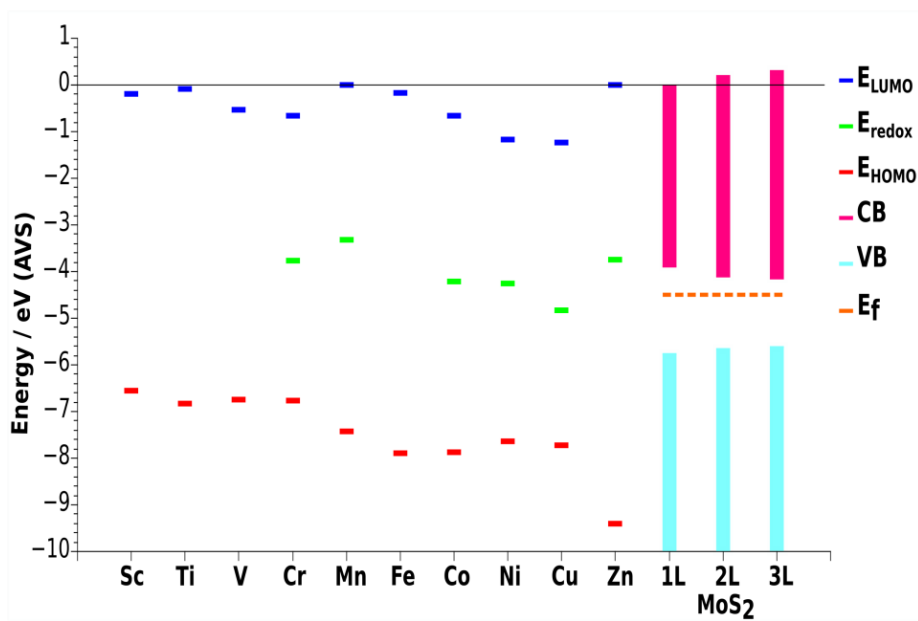

**Figure 8:** Diagram of HOMO and LUMO energy of the first row transition metals and energy values of electronic bands of $MoS_2$ in absolute vacuum scale. The red lines indicate HOMO energy of TM, blue lines indicate LUMO energy of TM, green lines indicate the redox potential of the TM pairs, light blue boxes represent the valence band of $MoS_2$, pink boxes represent the conduction band of $MoS_2$ and dot dashed orange line indicates the Fermi Level of $MoS_2$.

Due to the small energy difference between $MoS_2$ VB and CB, we can consider it a soft base (Absolute Hardness – $\eta$ ~ 0.9),[11] that will interact better with a soft acid. The fact that TM LUMO is relatively close to $MoS_2$ VB leads to a more covalent bond and a higher probability of a charge transference from $MoS_2$ to the TM. This behavior can be observed for Cr, Mn, Co, and Ni. This leads to the formation of $MoS_2$-TM coordination complexes with strong sigma covalent bonds, where S acts as a sigma donor ligand. Mn has an intermediary acid character ($\eta$ = 3.72) and induces a decrease in the electron density of $MoS_2$, shifting the Mo 3d peaks towards higher energies. Meanwhile, Cr is a softer acid ($\eta$ = 3.05) and tends to form a more covalent σ bond with S atoms, and the electrons on the $MoS_2$-Cr bond should be more delocalized. Additionally, S can act as π donor ligand, increasing the charge transfer to the TM and the observed blueshift on the XPS Mo spectra.

On the other hand, Co and Ni show a more donating character. Both are acid with intermediary hardness ($\eta$ of 3.6 and 3.2, respectively) and tend to form strong covalent bonds with S ligands due to their high ligand-field stabilization energy. Due to their filled 3d orbitals, Co and Ni complexes are known to form π back-bonds with soft ligands.[9] This way, for $MoS_2$-Ni and $MoS_2$-Co complexes, there is a σ donation from S to TM and a π back bonding involving the $3d_{xy, xz, yz}$ TM orbitals to empty orbitals on $MoS_2$ CB. The redshift observed on the Mo 3d core-level indicates that the charge transfer from Ni or Co to the $MoS_2$ is due to the more efficient back-donation. Additionally, this high degree of π back-bonding can increase the trion population explaining the observed trend on the exciton/trion ratio shown in Figure 7d. Since Co has smaller $E_{ads}$ than Ni (~ 1eV), a higher Co concentration is needed to reach a similar exciton/trion ratio (Figure S3).

However, Co and Ni present a very different behavior when analyzing the band gap shift in Figure 7c. $MoS_2$-Co samples presented a band gap shift of approximately 0.3 eV, while $MoS_2$-Ni samples showed no observable shift. Looking at the electronic band dispersion in Figure 2, we notice that $MoS_2$-$CoCl_3$ on the M site has a new state localized on Co atoms at ~0.25 eV above the CB. This new state can generate a $MoS_2 \rightarrow Co$ charge transfer transition with energy close to the band gap transition. Since our REELS measurement does not have enough resolution to distinguish both transitions, we observed an overall band gap reduction. For $MoS_2$-$CoCl_3$ on the H site or for $MoS_2$-Co (Figure S1), there are no new levels near CB, and no new transitions should be observed on the visible region. Thus, the band gap decrease observed for $MoS_2$-Co samples could indicate Co coordination on the M site and the Cl presence on its coordination sphere. For $MoS_2$-$NiCl_3$ complexes, the DFT calculations showed the existence of midgap states, and the electronic transition from the VB to these states should occur around 0.9 eV. This transition is close to the elastic peak on the REELS spectra and cannot be observed, which leads to no changes in the $MoS_2$ bandgap energy. We observe a very low signal in the Cl 2p spectra from the XPS of $MoS_2$-$NiCl_3$ and $MoS_2$-$CoCl_3$ (Figures S7 and S8). By comparing the peak area of Ni 2p or Co 2p peaks with the Cl 2p peaks as well as considering the relative atomic sensitivity factors (4.5 for Ni, 3.8 for Co and 0.73 for Cl) of XPS, we find that the overall atomic ratio of Ni:Cl and Co:Cl are close to 1:3.

Zn is the hardest TM in this series ($\eta = 4.72$), so there is a poor overlapping between its orbital and $MoS_2$'s ones, resulting in no variation of Mo 3d binding energy on the XPS spectra. Additionally, due to the Zn electronic configuration of $3d^{10}$, Zn complexes do not show ligand-field stabilization energy. Hardness and ligand-field stabilization energy both contribute to Zn's poor interaction with softer ligands, such as MoS$_2$, explaining the minimal BE energy shift and the slight decrease in the exciton/trion ratio.

For MoS$_2$-Cu complexes, by comparing Cu HOMO-LUMO levels with MoS$_2$ VB-CB energies, we expected a more covalent interaction, however, with a smaller E$_{ads}$, as shown in Figure 7, due to the Jahn-Teller effect[29]. However, looking at the TM's redox potential, we notice that Cu has a more negative value and it is the only one above the MoS$_2$ Fermi level. Thus, electrons on MoS$_2$ can be transferred to Cu spontaneously, reducing it from Cu$^{2+}$ to Cu$^{1+}$. A similar reduction effect was also reported in our previous paper that Au$^{3+}$ could be spontaneously reduced to Au$^{1+}$. It is worthy noting that we could not discard the possibility of Cu$^0$ formation; however, we did not observe Cu$^0$ in the XPS or the nanoparticle formation on SEM images. On the other hand, Cu$^{1+}$ does not have ligand-field stabilization energy due to its 3d$^{10}$ configuration. However, due to the absence of Jahn-Teller distortion and its softer character, the MoS$_2$-Cu bond energy should be higher than in Cu$^{2+}$ and Zn$^{2+}$ cases. These electron transfers from MoS$_2$ to Cu can lower the MoS$_2$ E$_f$, explaining the very high BE shift observed on Mo 3d XPS spectra. The high BE shift could also be pointing for the Mo$^{4+}$ oxidation into Mo$^{5+}$; however, Raman spectra of MoS$_2$-Cu samples (Figure S9) do not show any shift or new peaks compared to the pristine samples. Thus, we believe that Cu can induce a high degree of p-type doping on MoS$_2$, which can be supported by the decrease in trion population and the increase in the exciton/trion ratio, as shown in Figure 7d.

## Conclusion

In this work, we successfully prepared TM-MoS$_2$ coordination complexes for 3d transition metals Cr, Mn, Co, Ni, Cu, Zn and demonstrated that Ni forms single atoms that are bonded to S atoms via coordination bonds. This novel approach does not rely on defects and yet implies a significant impact on the optical and electrical properties of the functionalized monolayer MoS$_2$. The formation of the coordination complexes led to the transfer of electrons between MoS$_2$ and the transition metal, which can introduce both n- and p-type doping to MoS$_2$ depending on the transition metal used. Moreover, the degree of n and p-type doping can be fine-tuned by choosing the transition metal and varying the transition metal precursor concentrations, thus controlling the electronic band structure of MoS$_2$ and the exciton to trion relative population. In addition, the properties of the TM functionalized MoS$_2$ display a trend based on the 3d electron configuration of the transition metal that matches the periodic trend of well-studied coordination compounds. This trend can serve as a guide for future chemical functionalizations of monolayer TMD materials. The synthesis of single atoms introduced in this work could also be exploited in other applications such as single-atom catalysis (SAC), quantum information devices, optoelectronics, and enhanced sensing.

## Acknowledgments

This work was supported by the São Paulo State Foundation (FAPESP, grant nos. 2012/50259-8, 2015/11779-4 and 2017/01817-1), the Brazilian Nanocarbon Institute of Science and Technology (INCT/Nanocarbono), Conselho Nacional de Desenvolvimento Científico e Tecnológico (CNPq), and CAPES – PRINT (Programa Institucional de Internacionalização; Grant no. 88887.310281/2018-00). DG acknowledges financial support from Rio de Janeiro State Foundation (FAPERJ – grant nos. E-26/010.0022672019 and E-26/211.281/2019) and Serrapilheira Institute (grant number Serra – R- 2012-37959). This work was partly supported by the Air Force Office of Scientific Research (AFOSR) through grant No. FA9550-18-1-0072. LS acknowledges financial support from CNPq (Grant No. 408525/2018-5), and high-performance computing facilities from LoboC/NACAD/UFRJ.

**Author contributions:** The manuscript was written through contributions of all authors. H.L., W.C.S., and D.G. conducted the functionalization experiments and characterizations. L. S. G. S., A. G. V. and M. L. M. R. conducted the XPS and REELS measurements and analysis. L.S. performed the first-principle calculations. E.K. and K.T. synthesized the pristine MoS2 materials. K.F., T.Z. and F.Z. helped with TEM characterization and simulation. Y.L. and Z.Y. provided guidance and helped analyze the data. D.G., M.T., and C.J.S.M. supervised the whole work. All authors gave approval to the final version of the manuscript. H.L., W.C.S. contributed equally.

# Support Information

# Trends on 3d Transition Metal Coordination on Monolayer MoS$_2$

He Liu[1,†], Walner Costa Silva[2,†], Leonardo Santana Gonçalves de Souza[2], Amanda Garcez Veiga[2], Leandro Seixas[3], Kazunori Fujisawa[4,5,6], Ethan Kahn[7], Tianyi Zhang[7], Fu Zhang[7], Zhuohang Yu[7], Katherine Thompson[1], Yu Lei[7] Christiano J. S. de Matos[3], Maria Luiza M. Rocco[2], Mauricio Terrones[1,5,6,7,*], Daniel Grasseschi[2,*]

**Affiliations**
[1]Department of Chemistry, The Pennsylvania State University, University Park, PA, 16802.

[2]Institute of Chemistry, Federal University of Rio de Janeiro (UFRJ), 21941-909, Rio de Janeiro, Brazil.

[3]MackGraphe-Graphene and Nanomaterials Research Center, Mackenzie Presbyterian University, 01302-907, São Paulo, Brazil.

[4]Research Initiative for Supra-Materials (RISM), Shinshu University, 4-17-1 Wakasato, Nagano, 380-8553, Japan.

[5]Department of Physics, The Pennsylvania State University, University Park, PA, 16802.

[6]Center for 2-Dimensional and Layered Materials, The Pennsylvania State University, University Park, PA, 16802.

[7]Department of Materials Science and Engineering, The Pennsylvania State University, University Park, PA, 16802.

† - Equal contributors

* - Corresponding authors

**Table S1:** Calculated adsorption energy ($E_{ads}$), M-Mo distance and M-S bond length for first row transition metal coordination on surface

| Element | e Config | $E_{ads}$ (eV) | M-Mo (Å) | M-S (Å) |
|---|---|---|---|---|
| Sc | $3d^1 4s^2$ | -2.74 | 3.32 | 2.31 |
| Ti | $3d^2 4s^2$ | -3.02 | 2.99 | 2.32 |
| V  | $3s^3 4s^2$ | -2.88 | 2.84 | 2.28 |
| Cr | $3d^5 4s^1$ | -1.58 | 3.06 | 2.38 |
| Mn | $3d^5 4s^2$ | -1.41 | 2.96 | 2.35 |
| Fe | $3d^6 4s^2$ | -2.64 | 2.52 | 2.13 |
| Co | $3d^7 4s^2$ | -3.05 | 2.85 | 2.06 |
| Ni | $3d^8 4s^2$ | -3.85 | 2.58 | 2.13 |
| Cu | $3d^{10} 4s^1$ | -1.73 | 2.83 | 2.25 |
| Zn | $3d^{10} 4s^2$ | -0.40 | 4.13 | 3.16 |

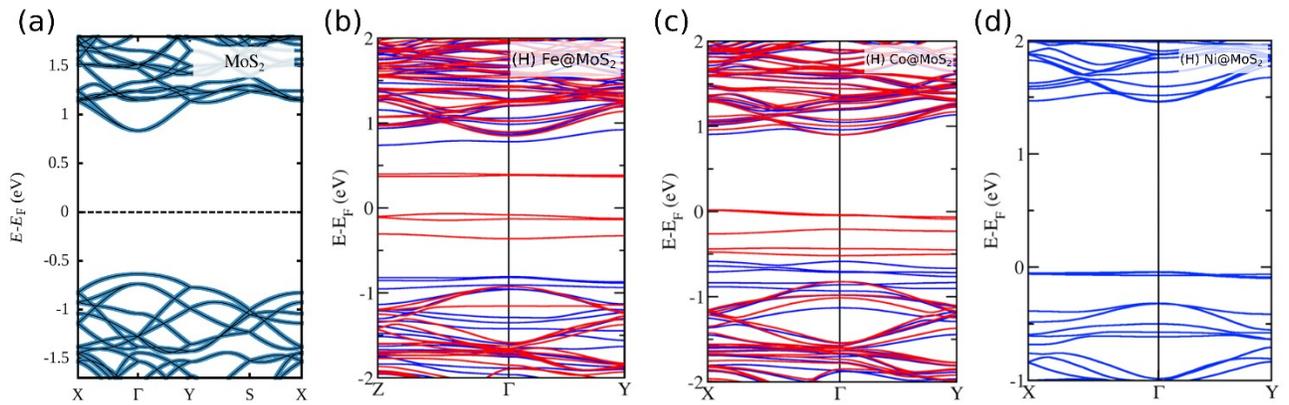

Figure S1: Electronic Band Dispersion of MoS2 (a) and Fe (b), Co (c) and Ni (d) functionalization on H site. The blue and red curves indicate states with different spin polarization.

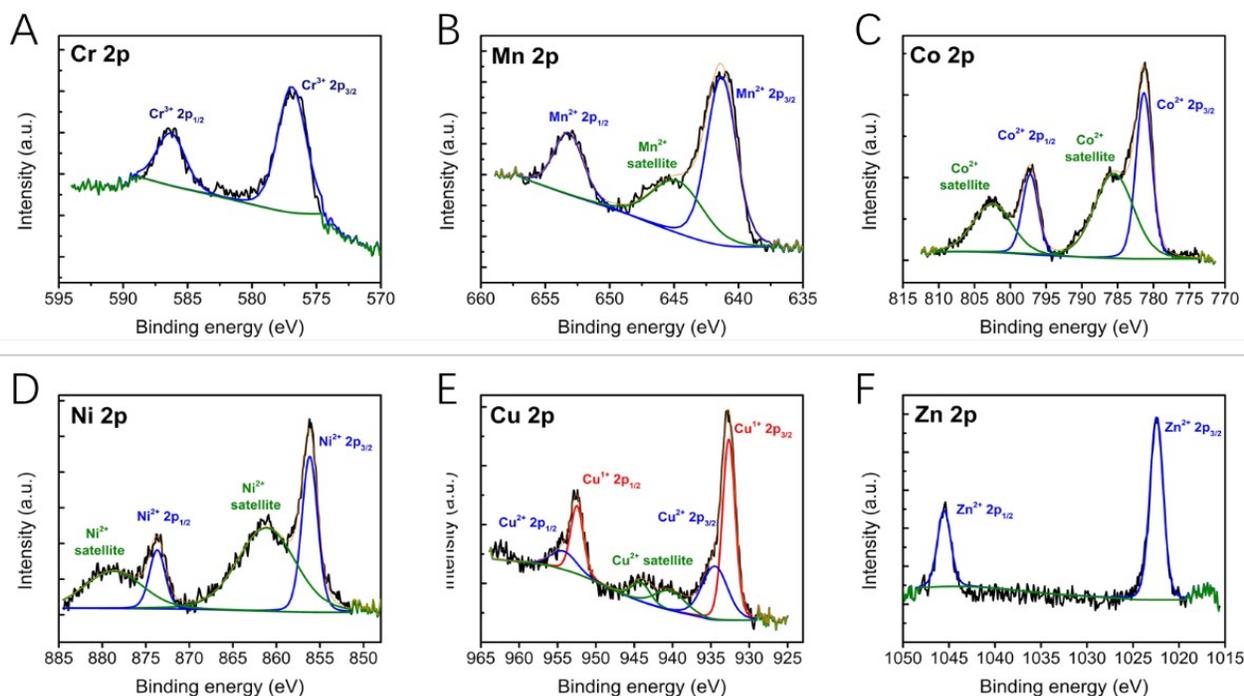

Figure S2: XPS spectra of TM 3d orbit after functionalization. (A) Cr 2p orbit of Cr-MoS$_2$. The black curve is the original spectra after C1s calibaration. The blue curves are fitted to Cr$^{3+}$ 2p$_{1/2}$ and 2p$_{3/2}$ peaks and they match perfectly with the original spectra. (B) Mn 2p orbit of Mn-MoS$_2$. The blue curves show the fitted Mn$^{2+}$ peaks. In addtion to the fitted Mn 2p$_{1/2}$ and 2p$_{3/2}$ peaks, a Mn satellite peak is also visible in the green curve. (C) and (D) Co and Ni 2p orbit of Co-MoS$_2$ and Ni-MoS$_2$. Similar satellite peaks could be fitted in addtion to the Co$^{2+}$ and Ni$^{2+}$ peaks. (E) Cu 2p orbit of Cu-MOS$_2$. Here the spectra are fitted into 3 groups. The Cu$^{1+}$ 2p peaks in red, the Cu$^{2+}$ 2p peaks in blue and the Cu$^{2+}$ satellite peaks in green. The presence of the Cu$^{1+}$ peaks indicate a reduction reaction between the Cu$^{2+}$ precusor and MoS$_2$. Similar to that of Au and MoS$_2$. (F) Zn 2p orbit of Zn-MoS$_2$ showing a perfect fit of Zn$^{2+}$ 2p$_{1/2}$, 2p$_{3/2}$ and the original spectrum.

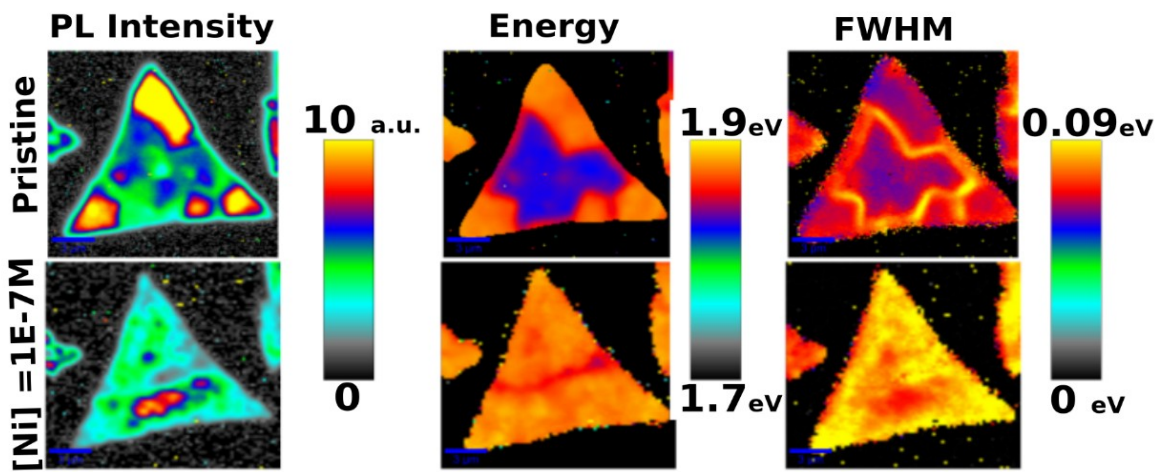

Figure S3: Photoluminescence mapping showing the PL intensity, energy of the maximum intensity, and the peak full width at half maximum (FWHM) of Pristine and Ni functionalized $MoS_2$. Intensity changes are uniform through the whole flake, thus indicating homogenous dispersion onto the flake's surface.

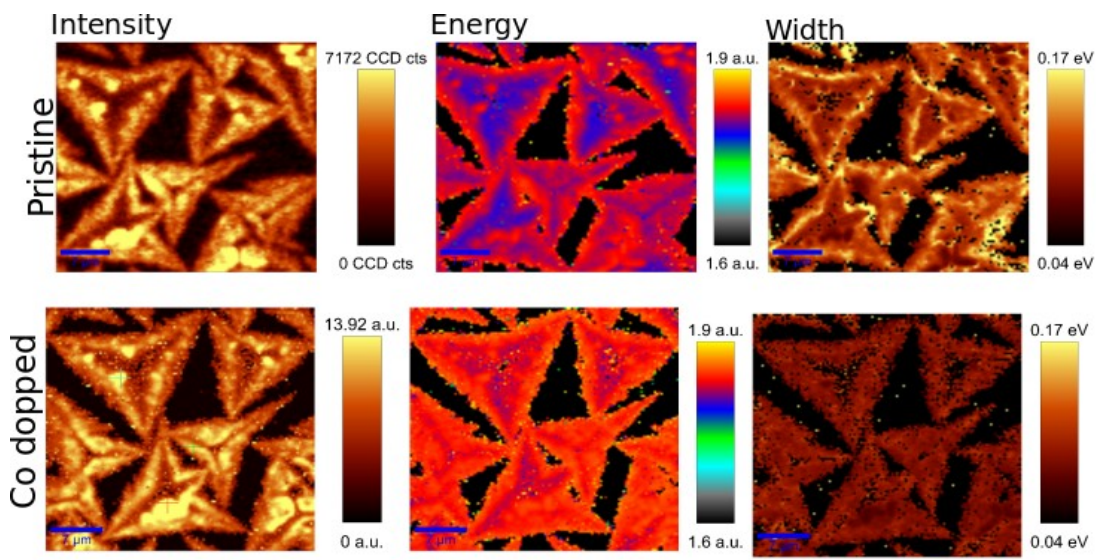

Figure S4: Photoluminescence mapping showing the PL intensity, energy of the maximum intensity, and the peak full width at half maximum (FWHM) of Pristine and Co functionalized MoS2. Intensity changes are uniform through the whole flake, thus indicating homogenous dispersion onto the flake's surface.

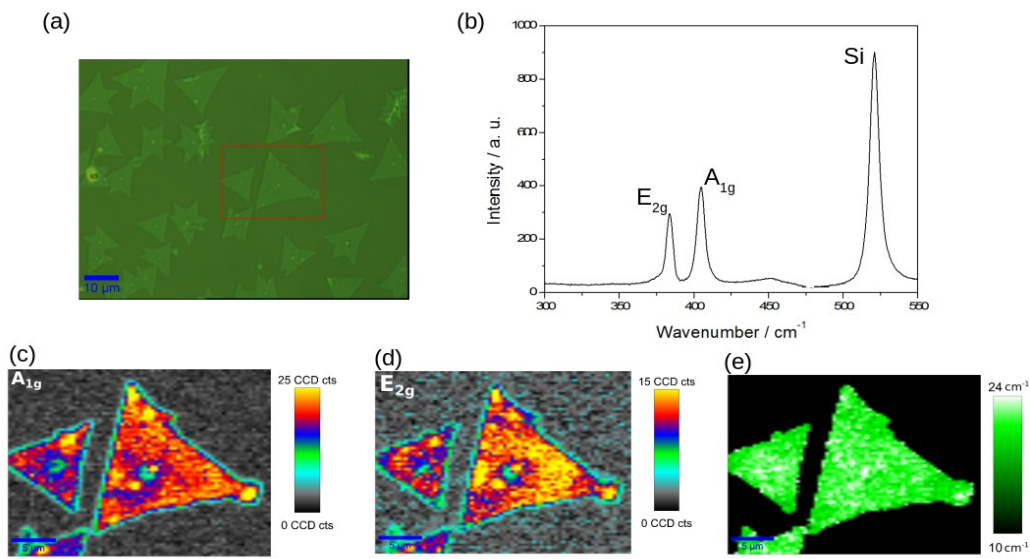

Figure S5: Optical microscopy of MoS$_2$-Ni samples (a). Raman spectra (b) and Raman mapping showing the A$_{1g}$ (c) and E$_{2g}$ (d) peaks intensity, and A$_{1g}$ and E$_{2g}$ wavenumber difference of MoS$_2$-Ni samples.

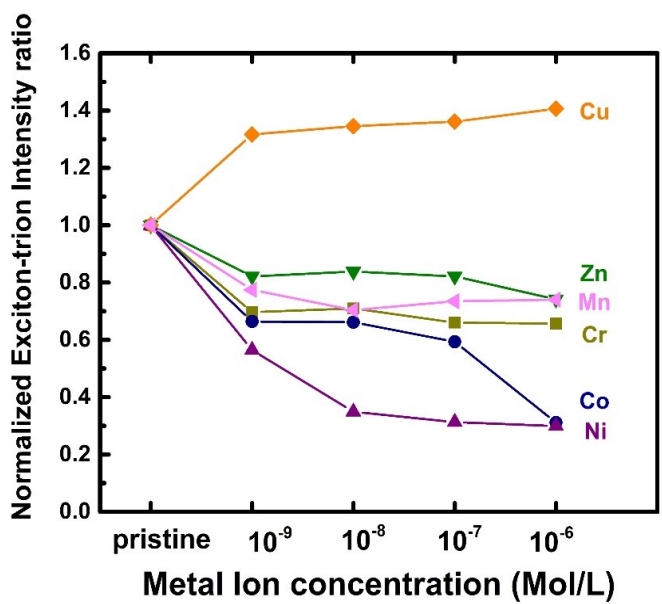

Figure S6: Normalized exciton to trion intensity changes showing the percent change in exciton to trion ratio for various TM functionalized MoS$_2$ samples at different TM concentrations. The TMs here can be separated into 3 groups. One is Cu, which is showing an increase to the exciton to trion ratio, indicating p-type doping on the MoS$_2$. The second group is Zn, Mn and Cr where we see a small decrease in the ratio, indicating weak n-type doping of MoS$_2$. The last group is Co and Ni where we see a large decrease in the ratio that corresponds to strong n-type doping of MoS$_2$.

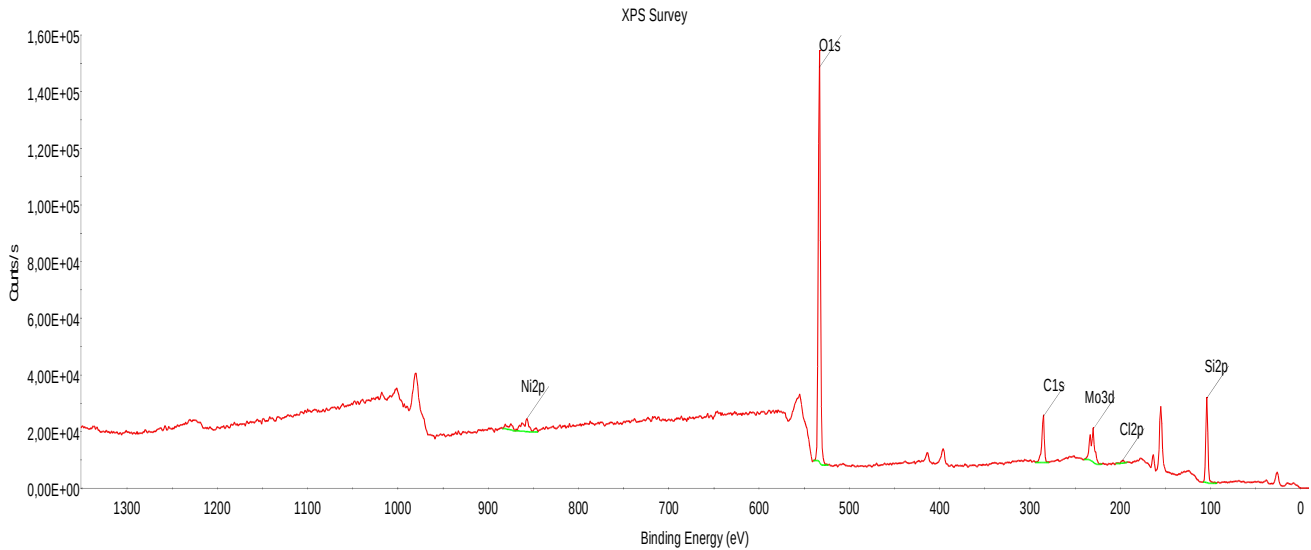

Figure S7: Survey XPS spectra of $MoS_2$-$NiCl_3$ samples showing the presence of MoS2 (Mo 3d and S 2p peaks), Ni (2p peak), and Cl (2p peak).

Table S2: Atomic percentage calculated from the $MoS_2$-$NiCl_3$ XPS survey spectra.

| Name | Peak BE | FWHM eV | Area (P) CPS.eV | Atomic % |
|---|---|---|---|---|
| O1s | 533.11 | 2.81 | 407121.28 | 49.32 |
| Si2p | 104.03 | 2.82 | 83899.53 | 29.76 |
| C1s | 285.22 | 2.9 | 54949.48 | 17.17 |
| Mo3d | 230.12 | 3.33 | 65903.51 | 2.12 |
| Ni2p | 857.04 | 3.56 | 55668.4 | 1.13 |
| Cl2p | 197.49 | 3.71 | 3831.59 | 0.5 |

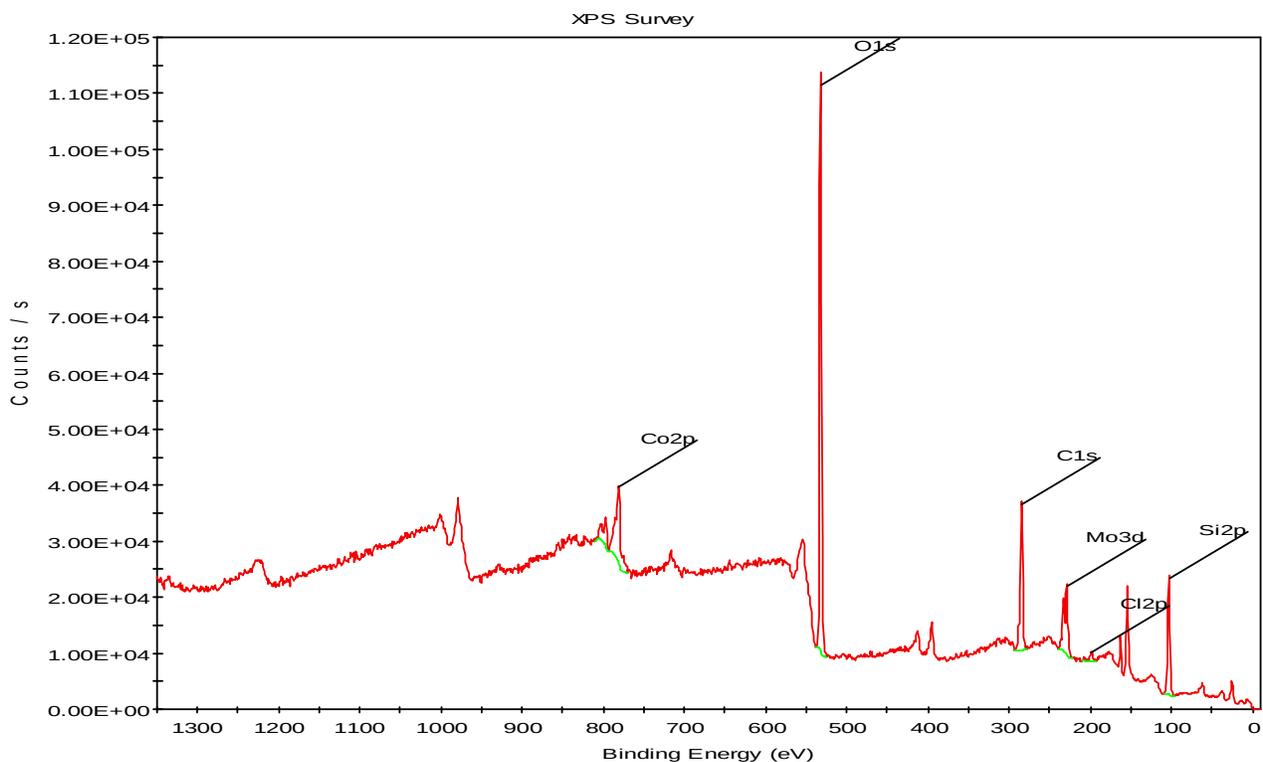

Figure S8: Survey XPS spectra of MoS$_2$-CoCl$_3$ samples showing the presence of MoS2 (Mo 3d and S 2p peaks), Co (2p peak), and Cl (2p peak).

Table S3: Atomic percentage calculated from the MoS$_2$-CoCl$_3$ XPS survey spectra.

| Name | Peak BE | FWHM eV | Area (P) CPS.eV | Atomic % |
|---|---|---|---|---|
| O1s | 532.1 | 3.002 | 315827.46 | 39.95 |
| Si2p | 103.09 | 2.937 | 63121.09 | 23.38 |
| C1s | 284.31 | 2.832 | 91342.13 | 29.79 |
| Co2p | 781.22 | 4.925 | 139550.87 | 3.21 |
| Mo3d | 229.21 | 3.207 | 72065.21 | 2.42 |
| Cl2p | 198.29 | 4.266 | 9115.36 | 1.25 |

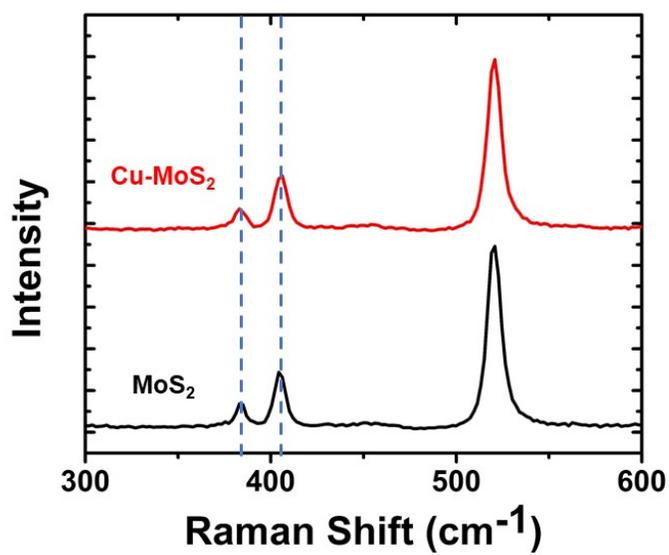

Figure S9: Raman Spectra of pristine and Cu functionalized MoS2. The MoS2 signature peaks do not change in wavenumber or intensity after Cu functionalization, indicating the functionalization process do not alter the intrinsic structure of the MoS2 lattice.